\documentclass[12pt,preprint]{aastex}
\newcommand{\kms}{km~s$^{-1}$}
\newcommand{\tpt}[1]{$\times 10^{#1}$}

\shorttitle{Probing the inner 200~AU of low-mass protostars with the SMA}
\shortauthors{J{\o}rgensen et al.}

\begin{document}

\title{Probing the inner 200~AU of low-mass protostars with the
  Submillimeter Array: Dust and organic molecules in NGC~1333-IRAS2A}

\author{Jes K. J{\o}rgensen\altaffilmark{1}, Tyler
L. Bourke\altaffilmark{1}, Philip C. Myers\altaffilmark{1}, Fredrik
L. Sch\"{o}ier\altaffilmark{2}, Ewine F. van Dishoeck\altaffilmark{3},
and David J. Wilner\altaffilmark{1}}
\altaffiltext{1}{Harvard-Smithsonian Center for Astrophysics, 60 Garden Street, MS42, Cambridge, MA 02138, USA ({\tt jjorgensen@cfa.harvard.edu})}
\altaffiltext{2}{Stockholm Observatory, AlbaNova, SE-106 91 Stockholm, Sweden}
\altaffiltext{3}{Leiden Observatory, PO Box 9513, NL-2300 RA Leiden, The Netherlands}

\begin{abstract}
The Submillimeter Array has opened a new window to study the innermost
warm and dense regions of the envelopes and disks around deeply
embedded protostars. This paper presents high-angular resolution
($<2$$''$) submillimeter observations of the class 0 young stellar
object NGC~1333-IRAS2A. Dust continuum emission and lines of complex
organic molecules such as CH$_3$OCH$_3$ and CH$_3$OCHO, high
excitation CH$_3$OH transitions, deuterated methanol CH$_3$OD as well
as lines of CO, HCN, HC$^{13}$CN, SO and SO$_2$ are detected on
$\lesssim 200$~AU scales. The observations are interpreted using
detailed radiative transfer models of the physical and chemical
structure, consistent with both single-dish and interferometer data.
The continuum emission is explained by an extended envelope and a
compact but resolved component, presumably a circumstellar disk with a
diameter of 200--300~AU and a mass of $\sim$ a few$\times
0.01-0.1$~$M_\odot$. If related to the rotation of the envelope, then
the size of this disk suggests a centrifugal barrier of 200--300 AU,
which implies that the temperature in the envelope does not increase
above 100~K. Its large size also suggests that the build-up of disks
proceeds rapidly throughout the early protostellar stages. The smaller
($<100$ AU) disks found around other deeply embedded protostars may be
a result of tidal truncation. The high-resolution observations of SO
can be explained with a simple constant abundance, $\sim$~$10^{-9}$,
constrained through single-dish observations, whereas those of
H$^{13}$CN and the organic species require high abundances, increased
by one to two orders of magnitude, or an additional compact source of
emission at small scales. The compact molecular emission could
originate in a hot core region of the inner envelope, but a more
likely reservoir is the circumstellar disk.
\end{abstract}


\section{Introduction}
Understanding the physical and chemical structure of the inner hundred
AU of the envelopes around protostellar objects is important because
part of this material will be included in circumstellar disks from
which planets may form eventually. The innermost envelopes are
characterized by high temperatures ($\sim$~100~K) and densities ($\sim
10^7-10^8$~cm$^{-3}$), making them eminently suited for observations
at submillimeter wavelengths.  These regions are, however, heavily
diluted in single-dish beams ($<2''$ size compared to typical
single-dish beam sizes of 10--20$''$) and interpretation of their line
and continuum emission relies on extrapolation of the density and
temperature distribution from observations on larger
scales. Furthermore, studies based on observations of lower excitation
lines are complicated by the fact that a large fraction of their
emission may arise in the outer cold regions of the envelopes, where
the lines can also become optically thick. Interferometry with the
Submillimeter Array (SMA) avoids most of these complications and thus
provides a unique possibility to probe the warm and dense material in
the innermost regions of protostellar systems.  In this paper we
present SMA observations for the low-mass protostar
NGC~1333-IRAS2A. This study will serve as an illustration of the type
of data and analysis techniques that are possible for these objects
and will be applied in the future to a much larger sample of low-mass
protostars currently being observed at the SMA.

The first goal of this project is to constrain the physical structure
on scales of a few hundred AU from the continuum data. The larger
($>1000$ AU) scale temperature and density distribution is well
determined from analysis of single-dish submillimeter continuum images
\citep[e.g.][]{shirley02,schoeier02,jorgensen02} but holes, cavities
and departures from spherical symmetry are likely at smaller radii.
In addition, any disk will start to contribute significantly to the
emission on scales of a few hundred AU. Since the dust continuum flux
scales with frequency as $\nu^2$ or steeper, submillimeter
interferometry is particularly well suited to probe the disks and
distinguish them from the envelope. Very little is known about the
physical properties such as mass and sizes of disks in these deeply
embedded stages, parameters which the SMA data can constrain.

A second goal of this project is to study the chemistry in the
innermost region, where large changes in abundances are expected due
to evaporation of icy grain mantles that inject species such as H$_2$O
and CH$_3$OH into the gas-phase. Reactions between evaporated
molecules result in a distinct ``hot core chemistry'' containing even
more complex organic molecules. In recent years the existence of such
hot cores has been suggested in a number of low-mass young stellar
objects \citep[e.g.,][]{maret04}, most strongly in the protostellar
binary IRAS~16293--2422 where warm gas and high abundances of a number
of species are thought to be present in the innermost ($r < 100$~AU)
regions \citep{vandishoeck95,ceccarelli00b,schoeier02}. Detections of
various complex organic molecules, e.g., HCOOH, CH$_3$OCH$_3$,
CH$_3$CHO, CH$_3$CN, and C$_2$H$_5$CN have been reported from
single-dish observations toward IRAS~16293--2422 \citep{cazaux03} and
another low-mass protostar NGC~1333-IRAS4A
\citep{bottinelli04n1333i4a}. Still, it is not clear how many low-mass
protostars show complex organic molecules and whether these reflect
the heating of ices or other processes such as the action of the
outflows \citep[see, e.g., discussion in][]{hotcoresample}. Yet
another fascinating question is whether any of the organic molecules
reside in the circumstellar disks. High angular resolution
interferometer observations of IRAS~16293-2422 for example
\citep{kuan04,bottinelli04iras16293} show that line emission of the
organic molecules peaks close to the position of two compact dust
continuum sources thought to mark the location of circumstellar disks
around each of the binary components.

This paper presents high angular resolution submillimeter observations
of dust continuum emission and high excitation transitions of organic
and other molecules toward the class 0 protostar NGC~1333-IRAS2A using
the SMA. The data are interpreted in the context of physical and
chemical models previously constrained from single-dish observations
and coupled with detailed radiative transfer calculations.  The paper
is laid out as follows: \S2 describes the details of the observations
and \S3 shows the qualitative results in terms of the detected
continuum and line emission - including the detections of high
excitation lines of, e.g., CH$_3$OH. \S4 presents a detailed analysis
based on the available continuum and line radiative transfer models
and \S5 discusses the origin of the observed emission.

\section{Observations}
NGC~1333-IRAS2A\footnote{In this paper we adopt a distance of 220~pc
  toward NGC~1333 \citep{cernis90}.} ($\alpha_{2000}=03^{\rm h}28^{\rm
  m}55\fs57$, $\delta_{2000}=+31^\circ14'37\farcs2$) \citep{looney00}
was observed on 2004 October 17 with the Submillimeter
Array\footnote{The Submillimeter Array is a joint project between the
  Smithsonian Astrophysical Observatory and the Academia Sinica
  Institute of Astronomy and Astrophysics, and is funded by the
  Smithsonian Institution and the Academia Sinica.}  \citep[][]{ho04}
at 350~GHz. The array had 7 antennae in the ``Compact-North''
configuration covering projected baselines from 18 to 164~k$\lambda$
corresponding to a beam size of
1.8\arcsec$\times$1.0\arcsec\ (P.A. 83$^\circ$) with natural
weighting. The weather conditions were excellent with typical 225~GHz
zenith sky opacities of 0.025--0.03. The 2 GHz bandwidth in each
sideband of the SMA digital correlator was configured with 1 chunk of
512 channels (345.853--345.957~GHz and 354.453--354.557~GHz) and 3
chunks of 256 channels (345.765--345.869, 345.279--345.383,
344.295--344.399~GHz and 354.541--354.645, 355.027--355.131,
356.011-356.115~GHz). With the width of each chunk of 104~MHz, the
resulting spectral resolutions are 0.2--0.4~MHz (0.15--0.30~\kms). The
remaining bandwidth was covered by 20 chunks each of 32 channels with
a resulting resolution of 3.25~MHz (2.8~\kms); no bright lines are
expected in these chunks, which are used to determine the
continuum. The separation between the upper and lower sideband is
10~GHz. The complex gains were calibrated through observations of the
two nearby quasars, 3C84 and 0359+509, the flux through observations
of Uranus and the passband through observations of Uranus and
Venus. The initial reduction was performed in the ``MIR'' package
\citep{qimir} and subsequently maps created and cleaned in Miriad
\citep{sault95}. The resulting RMS sensitivity was 15~mJy~beam$^{-1}$
over the 2~GHz bandwidth (continuum) and 0.2-0.35~Jy~beam$^{-1}$ per
channel in the narrow band setups.

\section{Results}\label{results}
\subsection{Continuum}
Continuum emission was clearly detected toward IRAS2A. The nearby
companion IRAS2B \citep{looney00,n1333i2art} is located about
30\arcsec\ away, i.e., outside the primary beam of the SMA 350~GHz
observations (half power beam width of 35\arcsec) and is not picked up
by the observations presented in this paper. A map of the continuum
emission is shown in Fig.~\ref{zerothorder}. For IRAS2A Gaussian fits
to the slightly extended emission (2.3$''$$\times$1.6$''$ FWHM) give a
total integrated flux of approximately 1~Jy (a peak flux of
0.46~Jy~beam$^{-1}$). For comparison, \cite{chandler00} report a flux
of 4.79$\pm $0.39~Jy at 850~$\mu$m in an aperture with a radius of
45$''$ from the JCMT/SCUBA whereas \cite{brown00} report a ``disk''
flux of $0.48^{+0.14}_{-0.13}$~Jy from interferometer observations
with the JCMT and CSO interferometer covering baselines of
$\approx$~70--190~k$\lambda$. These differences illustrate that a
large fraction of the emission is resolved out due to the
interferometer's lack of short spacings. In order to make any
(qualitative or quantitative) statements about the origin and
properties of the compact, emitting component, the extended envelope
therefore needs to be subtracted, and the interpretation of the
interferometric data requires a model.

\subsection{Lines}
\begin{figure}
\plotone{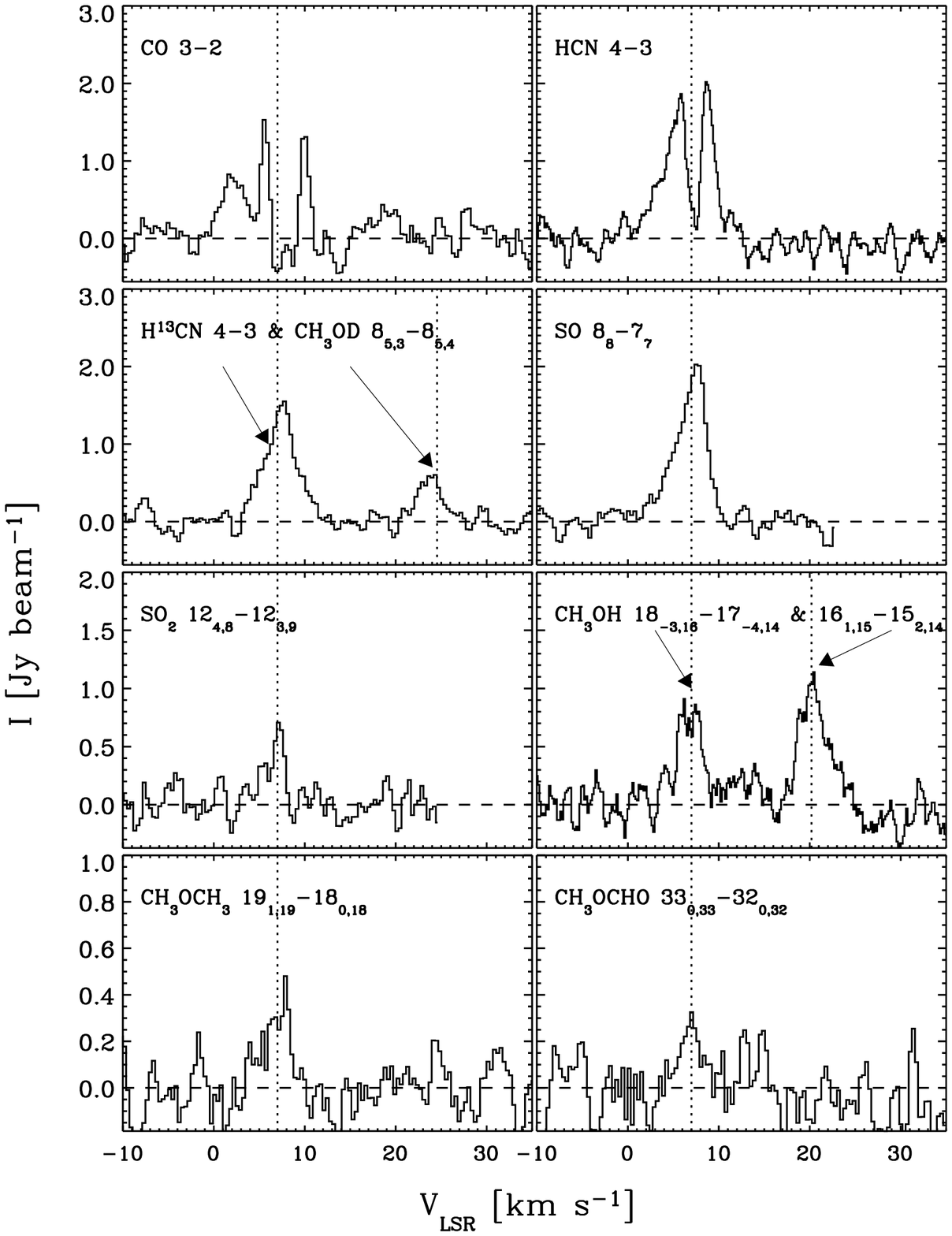}
\caption{Spectra of each observed species in the
  1.8\arcsec$\times$1.0\arcsec\ beam toward the (0,0) position. The
  dotted line indicates a systemic velocity of 7.0~\kms\ for each
  line.}\label{spectra_overview}
\end{figure}

In total 10 lines were detected at the 3$\sigma$ level in this single
track including the optically thick CO 3--2 and HCN 4--3 lines,
optically thin lines of H$^{13}$CN, SO, and SO$_2$ and high excitation
lines of organic molecules including CH$_3$OH, CH$_3$OD and
CH$_3$OCH$_3$ with a tentative detection of CH$_3$OCHO. The lines
detected are listed in Table~\ref{linelist} and their spectra in the
central beam shown in Fig.~\ref{spectra_overview}.
\begin{figure*}
\plotone{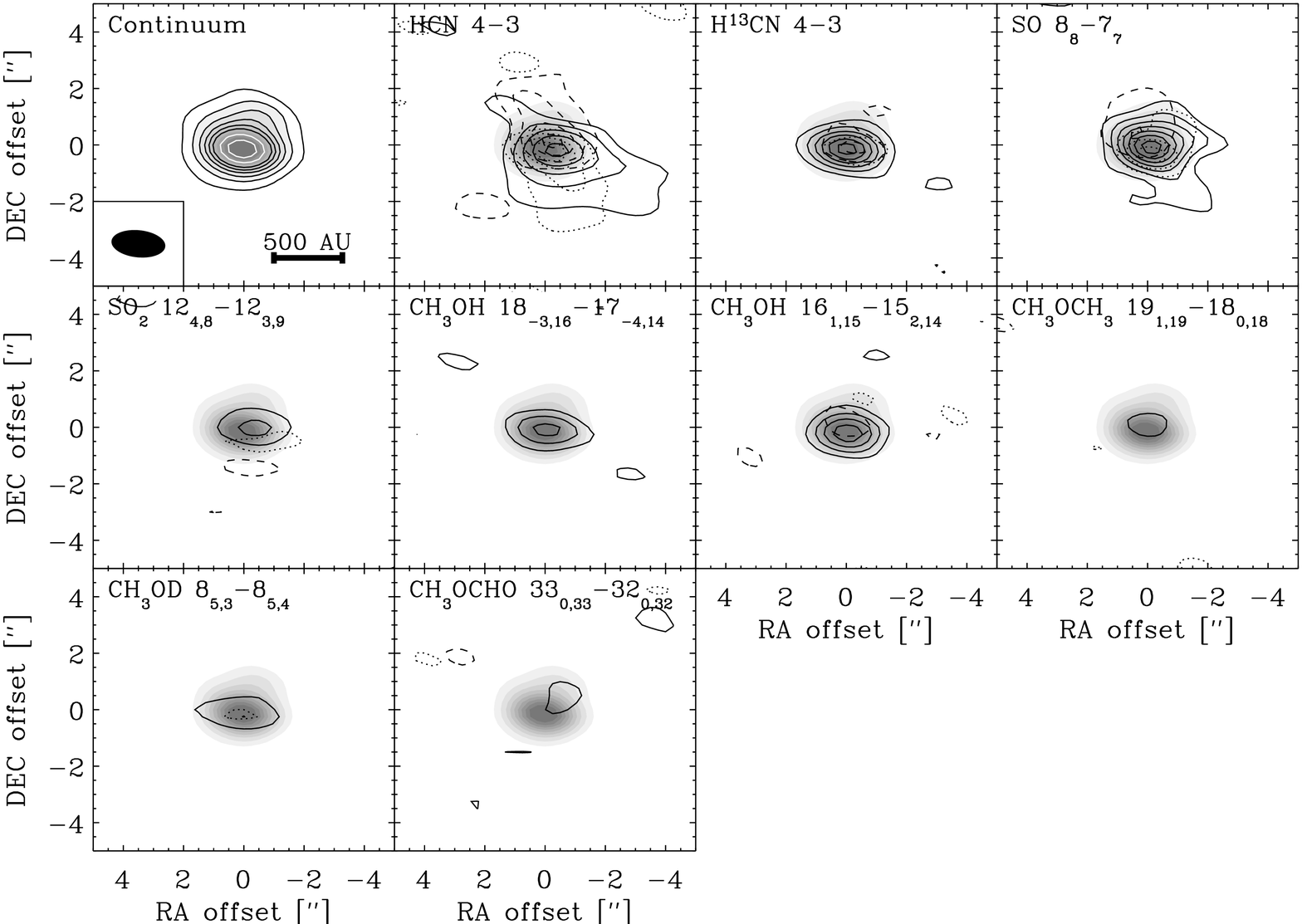}
\caption{Maps of the integrated emission of the continuum (upper left)
  and each of the molecular species, except CO 3--2. For each species
  the emission integrated over 2.5 to 5.5~\kms\ is shown with the
  dotted line, over 5.5 to 8.5~\kms\ with the solid line and over 8.5
  to 11.5~\kms\ with the dashed line. The contours are shown as black
  lines at 3$\sigma$, 6$\sigma$, $\ldots$, 18$\sigma$ (1$\sigma$ being
  15 mJy beam$^{-1}$ for the continuum and
  0.23~Jy~km~s$^{-1}$~beam$^{-1}$ for the line emission). In the
  continuum image, the white lines furthermore indicate the 24$\sigma$
  and 30$\sigma$ levels. The beam size is shown in the upper left
  panel together with a line indicating a 500~AU
  scale.}\label{zerothorder}
\end{figure*}

Results of Gaussian fits to the lines observed in the central beam
toward the continuum position are given in
Table~\ref{linefits}. Besides the CO 3--2 and HCN 4--3 lines that
clearly show structure due to resolved out extended emission at the
systemic velocity, and possible self-absorption, the lines are all
well represented by Gaussian profiles with line widths of
3--4~\kms\ (FWHM). SO$_2$ is the only exception with a narrower line
width of $\approx$~2~\kms.
\begin{deluxetable}{llll}
\tablecaption{Results of Gaussian fits toward the detected lines in the central
beam.\label{linefits}\label{linelist}}
\tablehead{\colhead{Line} & \colhead{Freq.} & \colhead{$I_{\rm peak}$\tablenotemark{a}}  & \colhead{$\Delta V$\tablenotemark{a}}  \\ \colhead{} & \colhead{[GHz]} & \colhead{[Jy beam$^{-1}$]}  & \colhead{[km s$^{-1}$]}}  \\
\startdata
CO 3--2                     & 345.7960         & $\ldots$\tablenotemark{b} & $\ldots$ \\
HCN 4--3                    & 354.5055         & $\ldots$\tablenotemark{b} & $\ldots$ \\
H$^{13}$CN 4--3             & 345.3398         & 1.4 (0.06)   & 4.2 (0.2)       \\
SO $8_8-7_7$                & 344.3108         & 1.9 (0.07)   & 3.9 (0.2)       \\
SO$_2$ $12_{4,8}-12_{3,9}$  & 354.0455         & 0.54 (0.07)  & 2.4 (0.4)       \\
CH$_3$OH $16_{1,15}-15_{2,14}$ (A-type)  & 345.9040    & 0.99 (0.07)  & 3.6 (0.3)       \\
CH$_3$OH $18_{-3,16}-17_{-4,14}$ (E-type)& 345.9192    & 0.79 (0.08)  & 3.4 (0.4)       \\
CH$_3$OD $8_{5,3}-8_{5,4}$ (A-type)  & 345.3196         & 0.58 (0.13)  & 3.1 (0.8)       \\
CH$_3$OCH$_3$ $19_{1,19}-18_{1,18}$ & 344.3579 & 0.32 (0.15)  & 3.5 (2.1)       \\
CH$_3$OCHO $33_{0,33}-32_{0,32}$ & 354.6078         & 0.27 (0.08)  & 2.8 (1.0)
\enddata

\tablenotetext{a}{Peak brightness and line width (FWHM) from Gaussian
  fits to the detected lines. The number in parenthesis indicates the
  estimated error from the fit.}  \tablenotetext{b}{These transitions
  show strong self-absorption and peak brightness and line width are
  not significant. Total emission integrated from $-$1 to 13~\kms\ is
  5.8~Jy~beam$^{-1}$~\kms\ and 10.1~Jy~beam$^{-1}$~\kms\ for the CO
  3--2 and HCN 4--3 lines, respectively.}
\end{deluxetable}

Maps of the observed line emission for selected species are shown in
Fig.~\ref{zerothorder}. Most of the species (except CO 3--2) show
compact emission around the continuum position and only CO, HCN,
H$^{13}$CN and SO are clearly resolved and show velocity gradients
associated with the outflow. The widths of the Gaussian fits to these
lines are not significantly larger than those of the remaining
species. The organic species and the SO$_2$ are largely unresolved
with the observing resolution corresponding to scales of
$200-400$~AU. This suggests that the observed emission either probes
the innermost region of the protostellar envelope (where the
temperatures are 50--60~K or higher) or is associated with the compact
component observed in the continuum data, possibly the circumstellar
disk.

\section{Detailed analysis}
\subsection{General considerations}
The development of detailed radiative transfer models has made it
possible to place constraints on the radial distribution of molecular
abundances throughout protostellar envelopes. In short, these models
use dust continuum observations to constrain the physical conditions
in the envelopes (e.g. temperature and density), which subsequently can
be used for non-LTE molecular excitation and line radiative transfer
calculations for comparison to both multi-line single-dish and
interferometer data. Interferometer observations of continuum emission
can be used to constrain the properties of the envelope on small
($\sim 100$~AU) scales and infer the presence of circumstellar disks
\citep[e.g,][]{harvey03,looney03} if emission from the envelope is
taken into account. Images of the continuum and line emission from the
protostar can be computed and then Fourier transformed for direct
comparison to the interferometer observations. This comparison in the
Fourier plane has two direct advantages: i) no image deconvolution and
restoration of the interferometer data is necessary and ii) the
comparisons between the models and interferometer data implicitly take
the problem of missing short spacings into account
\citep{n1333i2art}. The short spacings are included through the
single-dish observations used to constrain the physical and chemical
structure of the envelope in the first place.

As basis for a more detailed analysis we adopt the structure of the
NGC~1333-IRAS2 envelope derived from detailed dust and line radiative
transfer modeling \citep{jorgensen02, paperii,hotcoresample}. This
model assumes a spherically symmetric envelope heated by a central
source and the temperature profile is calculated self-consistently
through the dust radiative transfer. The single-dish observations are
well-fitted by a power-law density distribution, $n\propto r^{-p}$,
with $p=1.8$. The mass of the envelope is 1.7~$M_\odot$ within a
radius of 12,000~AU. This model has previously been compared to 3~mm
interferometer data from BIMA and OVRO and was found to explain well
the envelope structure down to scales of about 600-1000~AU
(3$''$-5$''$), the smallest scales probed \citep{n1333i2art}.

\subsection{Continuum emission: envelope and disk structure}\label{continstruct}
\begin{figure*}
\plottwo{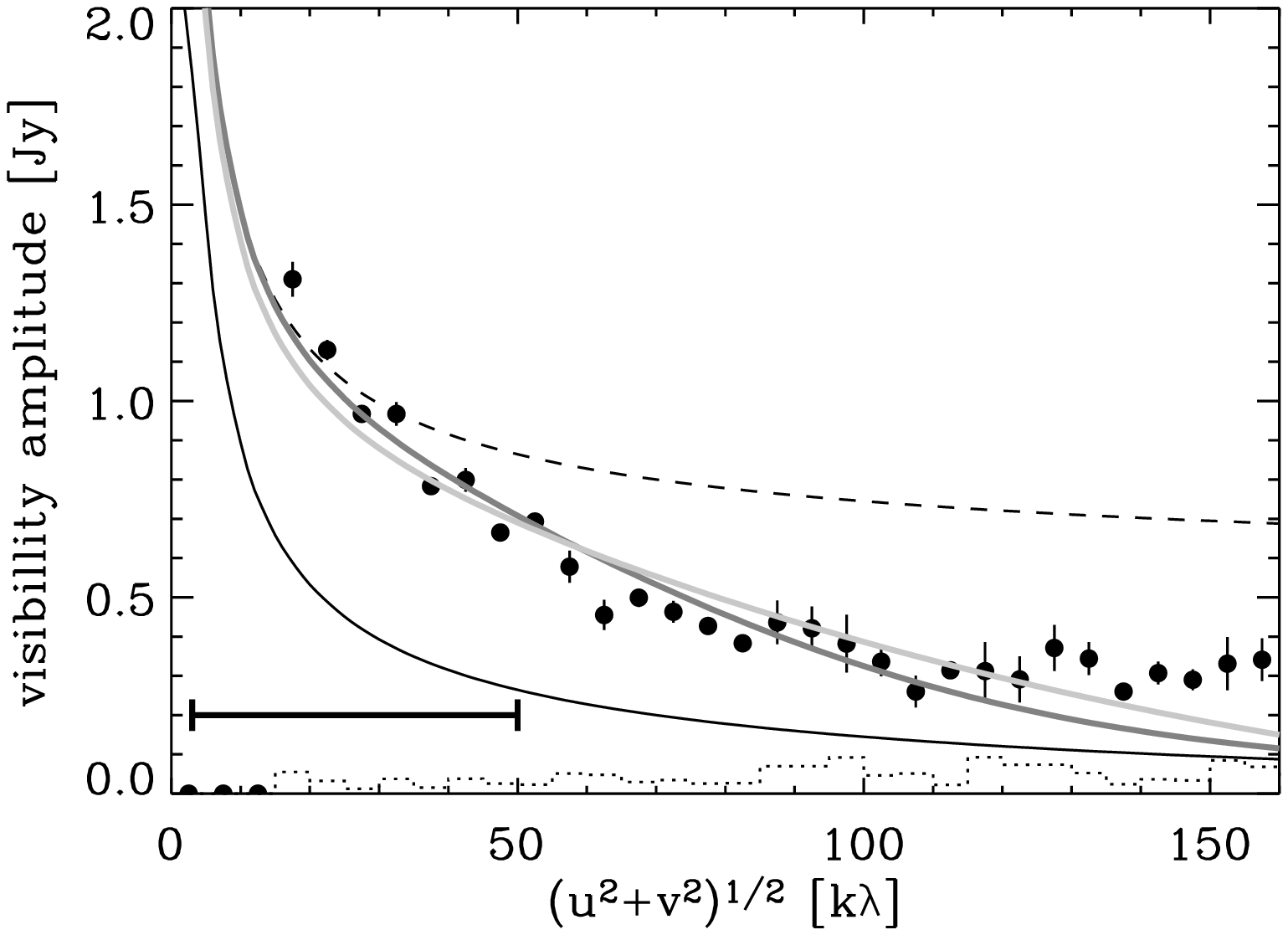}{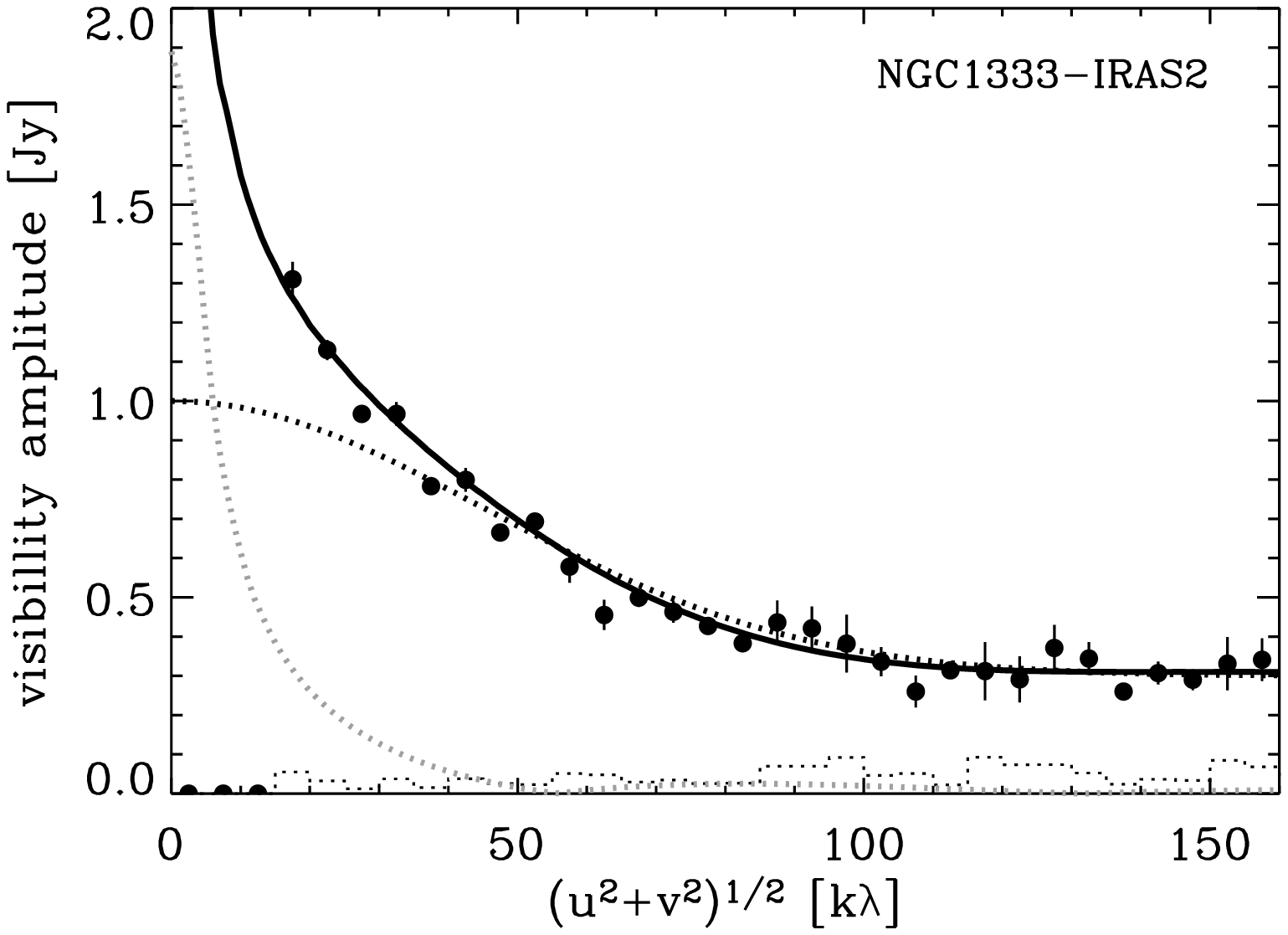}
\caption{Observed and modeled continuum visibility amplitude as
  function of projected baseline length. In both panels the error bars
  on each data-point are 1$\sigma$ statistical errors and the dotted
  histogram indicate the zero expectation level, i.e., the anticipated
  amplitude in the absence of source emission. In the left panel the
  solid black line indicates the envelope model of \cite{jorgensen02}
  based on single-dish data, the dashed black line the same model but
  with a constant 0.6~Jy added to all visibilities. The light-grey
  line is the original envelope model of \cite{jorgensen02} with an
  optimized ``Gaussian disk model'' with a flux of 0.6~Jy and a
  diameter of 1.1$''$ (250~AU). The dark grey line indicates a similar
  model but with an envelope with a cavity having a radius of 285~AU
  where the temperature has dropped to 75~K and a disk flux of 0.8~Jy
  and diameter of 0.9$''$ (200~AU). The horizontal black line at
  0.2~Jy indicates the projected baselines where continuum emission
  was detected in the OVRO and BIMA 3~mm observations of
  \cite{n1333i2art}. In the right panel the grey dotted line is an
  envelope model with an inner temperature of 75~K. The black dotted
  line is a ``two component disk'' model: a component with a Gaussian
  brightness distribution and a diameter of 300~AU and a flux of
  0.7~Jy and an unresolved (point) source with a flux of 0.3~Jy. The
  solid line indicates the combined model with the envelope and disk
  components.}\label{cont_uvamp}
\end{figure*}

Fig.~\ref{cont_uvamp} compares the observed visibilities from the SMA
observations with the envelope models \citep{jorgensen02}. Consistent
with the 3~mm interferometer observations \citep{n1333i2art} a compact
continuum source is required to explain the observed emission in
addition to the extended envelope picked up at the shorter
baselines. In the 3~mm observations continuum emission was seen out to
baselines of $\approx 50$~k$\lambda$ which did not resolve the central
source. At the longer baselines observed with the SMA the visibility
amplitudes continue to decline indicating that the compact component
is resolved at scales of about 1$''$ ($\approx 220$~AU). This is
consistent with the result of \cite{brown00} who suggested that the
JCMT-CSO observations with baselines of $\approx$~70--190~k$\lambda$
resolved the disk around IRAS2A.

To constrain the properties of this central component, the data were
fitted with the prediction from the envelope model explaining the
larger scale emission but adding a Gaussian brightness distribution
simulating the disk. The Gaussian brightness distribution was taken to
be spherically symmetric and its best fit flux and width constrained
through estimates of the $\chi^2$ statistics. For an inner radius of
the envelope of 23~AU (corresponding to a temperature of 250~K), a
Gaussian distribution with a FWHM of 1.1$''$ (a diameter of $\approx
250$~AU) and a total flux of 0.6~Jy provides the best fit. This is
somewhat larger than the size inferred by \cite{brown00} of $\approx
100$~AU (scaled to a distance of 220~pc). This difference reflects i)
the shorter baselines included in the observations in this paper
constraining the disk contribution on larger scales and ii) the fact
that \citeauthor{brown00} did not include the contribution from the
envelope in their estimates. If an inner cavity exists in the envelope
such as suggested for IRAS~16293--2422 \citep[][J{\o}rgensen et al.,
  in prep.]{hotcorepaper} the flux of the compact component would be
higher. For example, for a cavity with a radius of $\approx$~285~AU
(corresponding to an inner temperature of 75~K) the flux of the
central component increases to 0.8~Jy and its diameter decreases to
0.9$''$ ($\approx 200$~AU). Note that this is just an example: the
interferometer does not place exact constraints on the cavity size and
therefore some degeneracy exists between the envelope and disk
parameters. Observations of the mid-infrared source with the Spitzer
Space Telescope would place exact constraints on the dust column
density toward the source center and could thereby resolve this
degeneracy.

No matter which inner radius is assumed for the model (23~AU or
285~AU), the flux at the longest baselines is underestimated. Varying
the slope of the density profile within the uncertainties allowed by
the SCUBA maps does not improve the best fit at the longer baselines
as the emission there is dominated by the Gaussian disk
component. Naturally the Gaussian profile is probably an
oversimplification of the disk structure on the smallest scales. This
point is illustrated by the solid line in the rightmost panel of
Fig.~\ref{cont_uvamp} which includes an envelope (with an inner cavity
of radius 285~AU), a disk with Gaussian brightness distribution (flux
of 0.5~Jy and diameter of 300~AU) and an unresolved 0.3~Jy point
source component.  This two component model improves the fit
statistically -- even taking the extra degree of freedom, introduced
with the unknown point source flux, into account. \cite{mundy96}
reached a similar conclusion for the disk around the T Tauri star HL
Tau: 2.7~mm and 0.87~mm continuum emission toward this source was
found to be more centrally peaked than a single Gaussian.
\citeauthor{mundy96} attributed this central condensation in emission
to temperature and density gradients in a simple power-law disk model.

What this discussion emphasizes is the need for good constraints on
the continuum emission on a wide variety of interferometer baselines
and from single-dish observations coupled with reliable models for
both the envelope and disk components. Thereby it becomes possible to
simultaneously constrain the flux of both the extended envelope, and
of the disk on intermediate (few hundred AU) and small scales
($<100$~AU) but naturally these estimates rely on our assumptions
about the disk and envelope structure.

\subsection{Line emission: abundance structure}\label{lineabund}
The line observations are compared to the predictions based on models
of the envelope constrained by the single-dish observations of
\cite{paperii,hotcoresample}. In those papers abundances of
H$^{13}$CN, SO and CH$_3$OH were determined and an upper limit of the
SO$_2$ line reported. In this paper, we focus on the chemistry of the
optically thin species -- and defer the discussion of CO and HCN to a
future paper. The aim is to constrain the abundance profiles in the
innermost regions which were difficult to probe with single-dish
telescopes or low-frequency interferometers. For H$^{13}$CN, SO and
SO$_2$ collisional rate coefficients from the \emph{Leiden Atomic and
  Molecular Database} \citep{schoeier03radex} were used to calculate
the NLTE molecular excitation. For the remaining species no data exist
for the observed high excitation transitions and LTE was assumed. For
CH$_3$OH one of the observed transitions probes A-type and the other
E-type (the distinction reflecting the rotation of the OH group with
respect to the CH$_3$ group), and the abundance of CH$_3$OH quoted in
this paper refers to the sum of the abundance of these two
species. The abundance structures for all species are summarized in
Table~\ref{abundstruct}. Each model is consistent with both
single-dish and interferometer observations with $\chi^2_{\rm red}
\lesssim 2$ - except for CH$_3$OH for which the single-dish data
indicate a hot core abundance an order of magnitude lower.

\subsubsection{SO and SO$_2$}
Fig.~\ref{so_uvplot}a compares the model prediction for the SO
$8_8-7_7$ from \cite{paperii} with the SMA results. A good fit is
obtained with a constant abundance of $\approx$~2.5--3$\times 10^{-9}$
in agreement with the results of \cite{paperii}. It is noteworthy how
well the SMA data are described by this constant abundance model. The
single-dish observations of \cite{paperii} are not sensitive to the
abundance in the innermost ($T>90$~K) region. In IRAS~16293--2422
\citep{schoeier02} an enhancement by two orders of magnitude in the
innermost, $T > 90$~K, region was inferred on the basis of single-dish
observations. In the case of IRAS2A, such enhancements appear to be
ruled out by the interferometer observations presented in this paper.

In contrast to SO, SO$_2$ only shows unresolved emission close to the
central protostar. \cite{paperii} did not detect SO$_2$ in the
NGC~1333-IRAS2 envelope and the detection in this paper suggests a
compact origin of the emission from this species with a jump in
abundance of at least two orders of magnitude. The absence of a clear
SO abundance enhancement at small scales is interesting in this
context. Simply comparing our observed [SO$_2$]/[SO] ratio $\gtrsim
2.5$ to the models of the sulphur chemistry by \cite{wakelam04hotcore}
would indicate that NGC~1333-IRAS2 is more evolved than
IRAS~16293--2422. In terms of absolute abundances, however, the upper
limit of $8\times 10^{-9}$ of SO is in contradiction with any of the
hot core models that predict SO abundances $\gtrsim 10^{-7}$
\citep[e.g.][]{charnley97,wakelam04hotcore}. These results therefore
suggest that SO does not probe a hot core around IRAS2A - or that its
chemistry is significantly different from other known hot cores.

\subsubsection{H$^{13}$CN}
For H$^{13}$CN no good fit to the single-dish observations is obtained
with a constant abundance model. This is similar to the case for CO
and HCO$^+$ where constant abundance models underestimate the
intensity of the lowest excitation 1--0 transitions. \cite{coevollet}
suggested that the abundances vary radially with a so-called ``drop''
profile: in these models freeze-out occurs in the region of the
envelope where the temperature is low enough to prevent immediate
desorption, but where the density is high enough that the freeze-out
timescales are shorter than the lifetime of the core. Similar type
abundance profiles have been confirmed by high angular resolution
observations of H$_2$CO and CO toward other protostars
\citep{hotcorepaper,l483art} and found in detailed chemical-dynamical
models \citep{lee04}. For the H$^{13}$CN 4--3 observations in this
paper (Fig.~\ref{h13cn_uvplot}b) and the single-dish observations
\citep{paperii} a combined best fit is obtained with an abundance
profile where depletion occurs at the same radius as it does for CO
and HCO$^+$ at a density of $7\times 10^{4}$~cm$^{-3}$ and evaporation
at small radii unresolved by the interferometer observations where the
temperatures exceeds 70--90~K (radii smaller than
$\approx$~100~AU). This is expected if HCN comes off dust grains at
roughly the same temperatures as H$_2$O ice mantles \citep{doty04}. In
terms of tracing the dense gas in the envelope, HCN and H$^{13}$CN
appear to be most useful probes of the inner envelope and/or disk with
high abundances, strong transitions and only weakly affected by
outflows (as it is also the case for the lower excitation $J=1-0$
lines \citep[e.g.,][]{n1333i2art}).

\subsubsection{CH$_3$OH and CH$_3$OCH$_3$} 
\begin{figure*}
\begin{center}
\plottwo{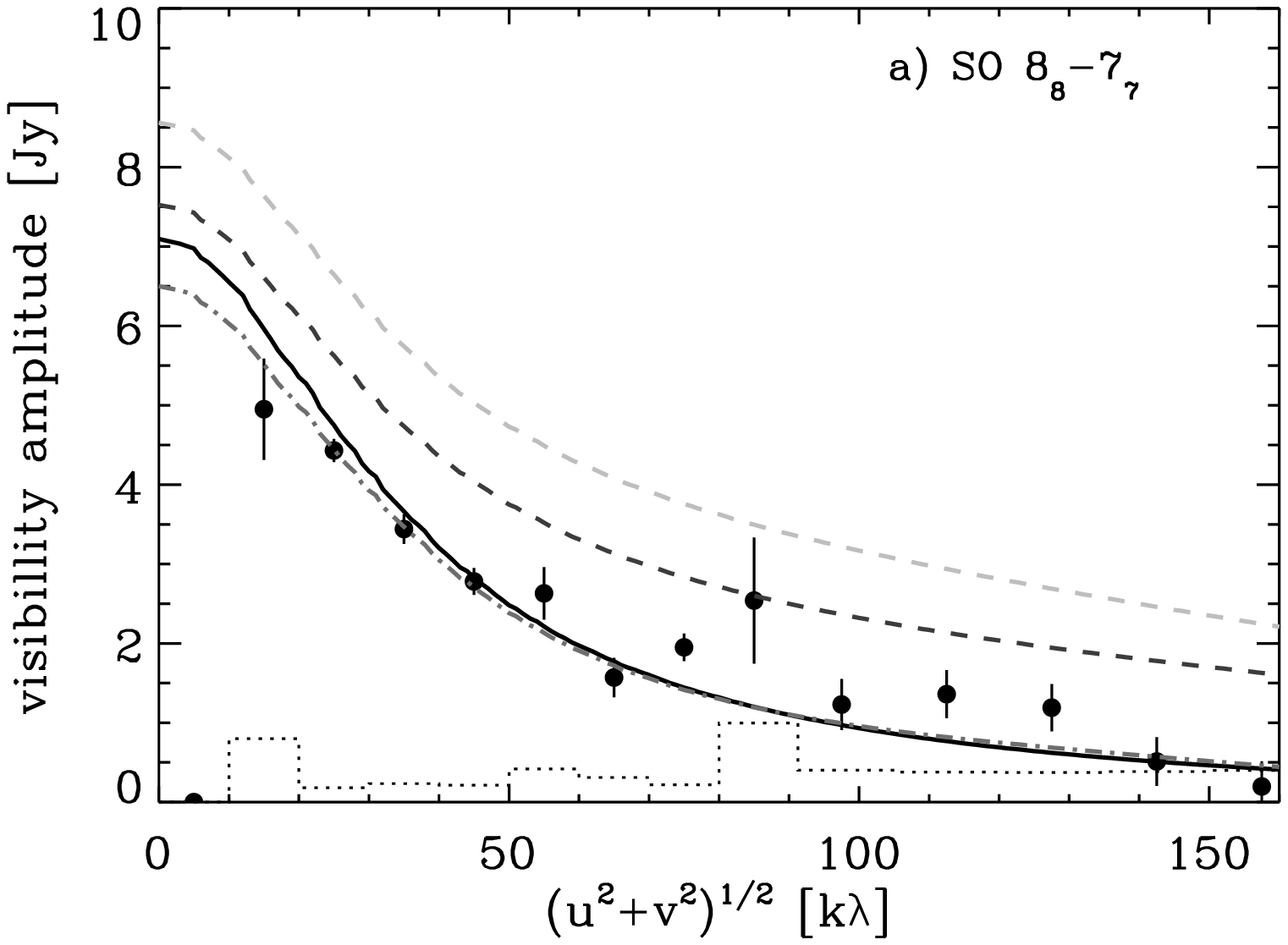}{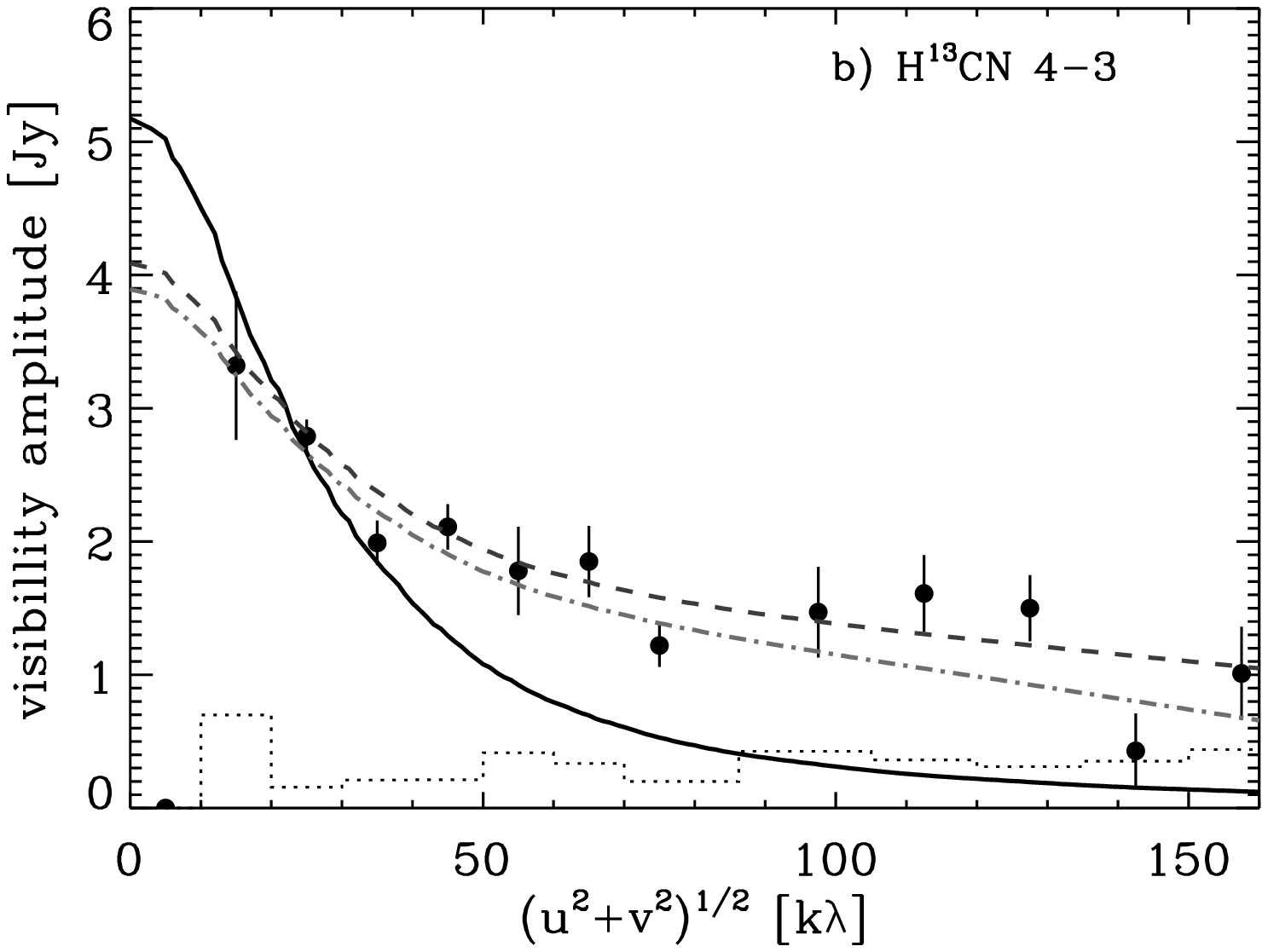}
\plottwo{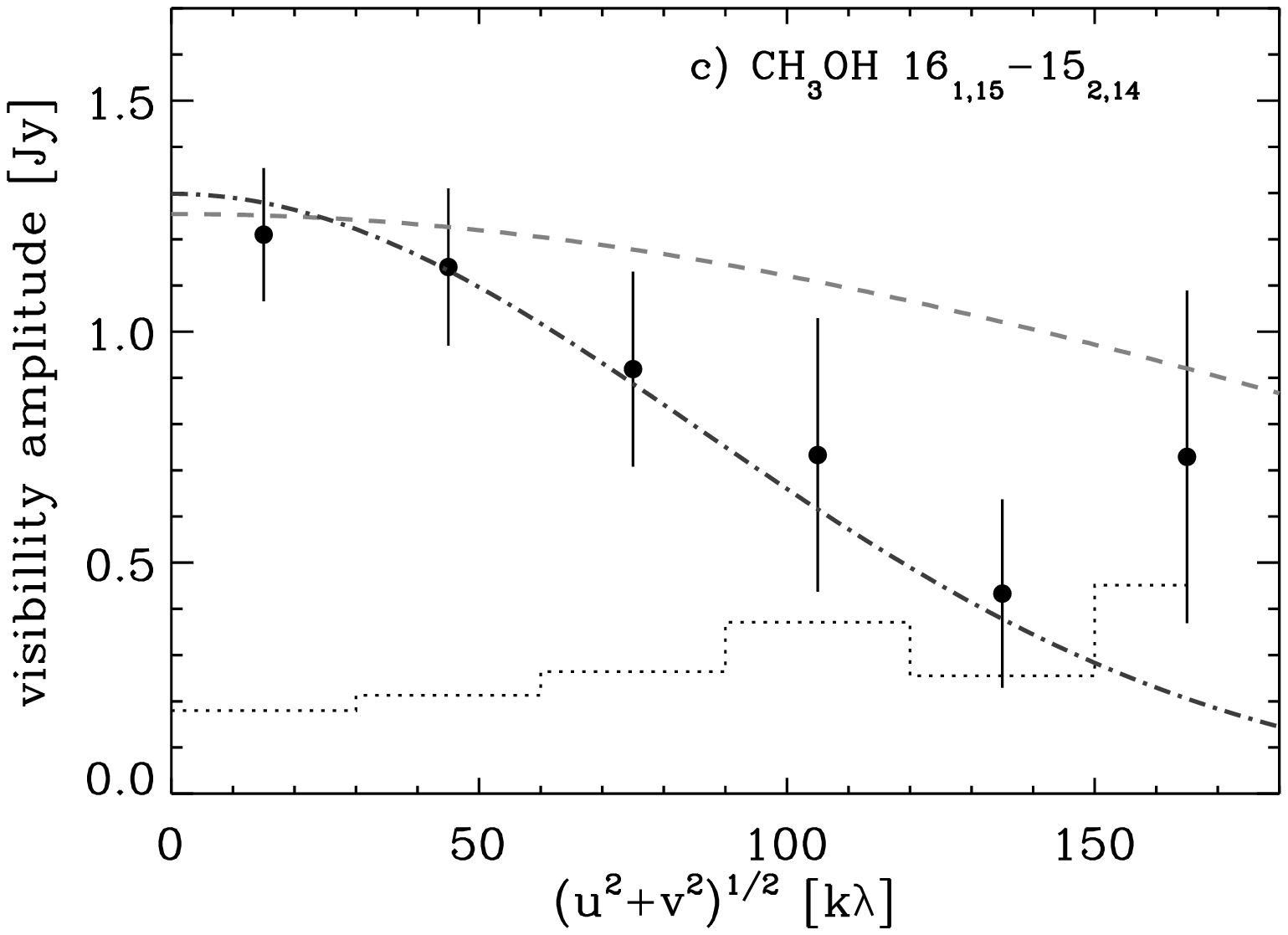}{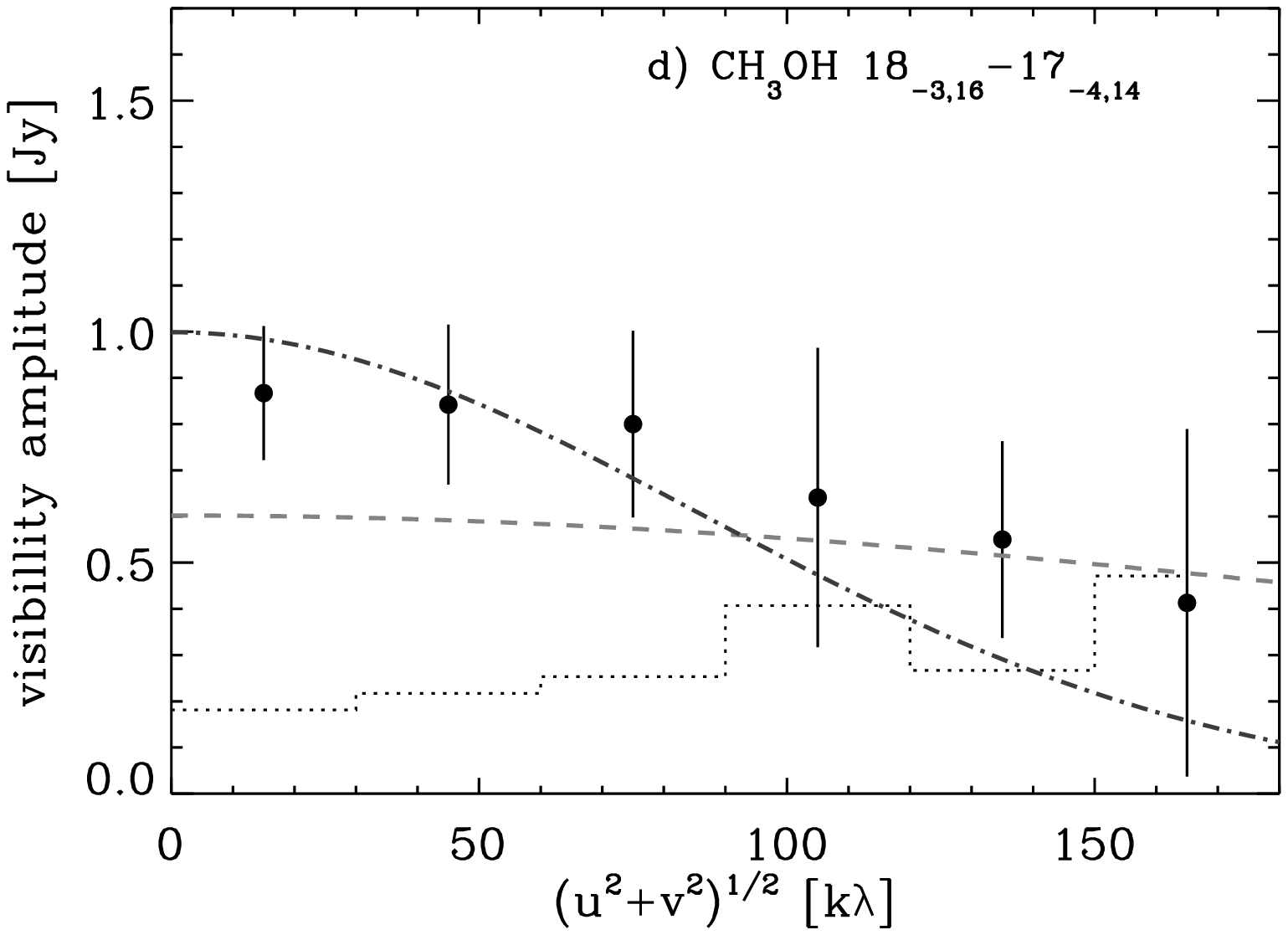}
\end{center}
\caption{Comparison between SMA observations and various abundance
  models for SO (a), H$^{13}$CN (b) and the CH$_3$OH
  $16_{1,15}-15_{2,14}$ (c) and $18_{-3,16}-17_{-4,14}$ (d)
  transitions. For each molecule the visibility amplitudes are
  averaged over channels covering velocities from 5 to 9~\kms. The
  error bars indicate 1$\sigma$ errors and the dotted histogram the
  zero-expectation level (i.e., the anticipated amplitude in the
  absence of source emission). In each panel the solid line indicates
  the best fit model for the given species from
  \cite{paperii,hotcoresample} (constant abundance for SO and
  H$^{13}$CN and jump abundance for CH$_3$OH) and the dashed-dotted
  line the envelope model with a cavity and a disk with a Gaussian
  brightness distribution (FWHM of 200~AU) and abundance given in
  Table~\ref{abundstruct}. For CH$_3$OH, the model from the
  single-dish data does not produce any measurable flux and is not
  plotted. The dashed lines indicate additional models: in panel (a)
  models with SO abundance enhancements in the inner regions by
  factors 10 (dark) and 100 (light). In panel (b) model with
  H$^{13}$CN drop abundance profile (Table~\ref{abundstruct}) and in
  panels (c) and (d) models with hot core abundances of $2\times
  10^{-6}$ for
  CH$_3$OH.}\label{so_uvplot}\label{h13cn_uvplot}\label{ch3oh_uvplot}
\end{figure*}

For the organic molecules the emission appears compact and mostly
unresolved. For these species the abundances in the region where $T >
90$~K (diameter $<130$~AU or scales of 0.6$''$; 300~k$\lambda$) are
constrained by the observed high excitation lines, whereas their line
intensities are largely independent of the abundance in the region
outside this radius. For CH$_3$OH the abundance in the inner envelope
derived from the interferometer data is an order of magnitude higher
than that derived on the basis of multi-transition single-dish
observations \citep{hotcoresample}. The CH$_3$OH observations shown in
Figs.~\ref{ch3oh_uvplot}c and \ref{ch3oh_uvplot}d do show a slowly
decreasing visibility amplitude with increasing baseline lengths, but
the data are consistent with an unresolved component at the 2$\sigma$
level (dashed lines in Fig.~\ref{ch3oh_uvplot}c and d). An even better
fit is obtained with a slightly extended ($\approx$150-200~AU)
Gaussian component (dashed-dotted lines).

\section{Discussion}
\subsection{Origin of the compact structure}\label{continorigin}
As explained above and seen for other low-mass protostars an
additional source of continuum emission has to be present in addition
to the envelope to explain the interferometer data. The SMA observes
the brightness distribution of this source and can thereby constrain
its physical parameters.

Taking the compact flux from the continuum data and adopting a dust
opacity of 1.82~cm$^{2}$~g$^{-1}$ at 850~$\mu$m \citep[][used in the
  envelope models]{ossenkopf94}, the mass (gas+dust) of a disk with a
temperature of 150~K is 0.02~$M_\odot$. This in turn corresponds to an
average H$_2$ column density of 1.6$\times 10^{24}$~cm$^{-2}$ for a
circular disk filling the beam with a diameter of 200~AU. Even in
models where the envelope does not have an inner cavity, a disk column
density of $\sim 10^{24}$~cm$^{-2}$ is non-negligible compared to the
hot core column density (i.e., pencil beam column density of the
material in the envelope with a temperature higher than 90~K) of
3$\times 10^{23}$~cm$^{-2}$.

\begin{figure}
\plotone{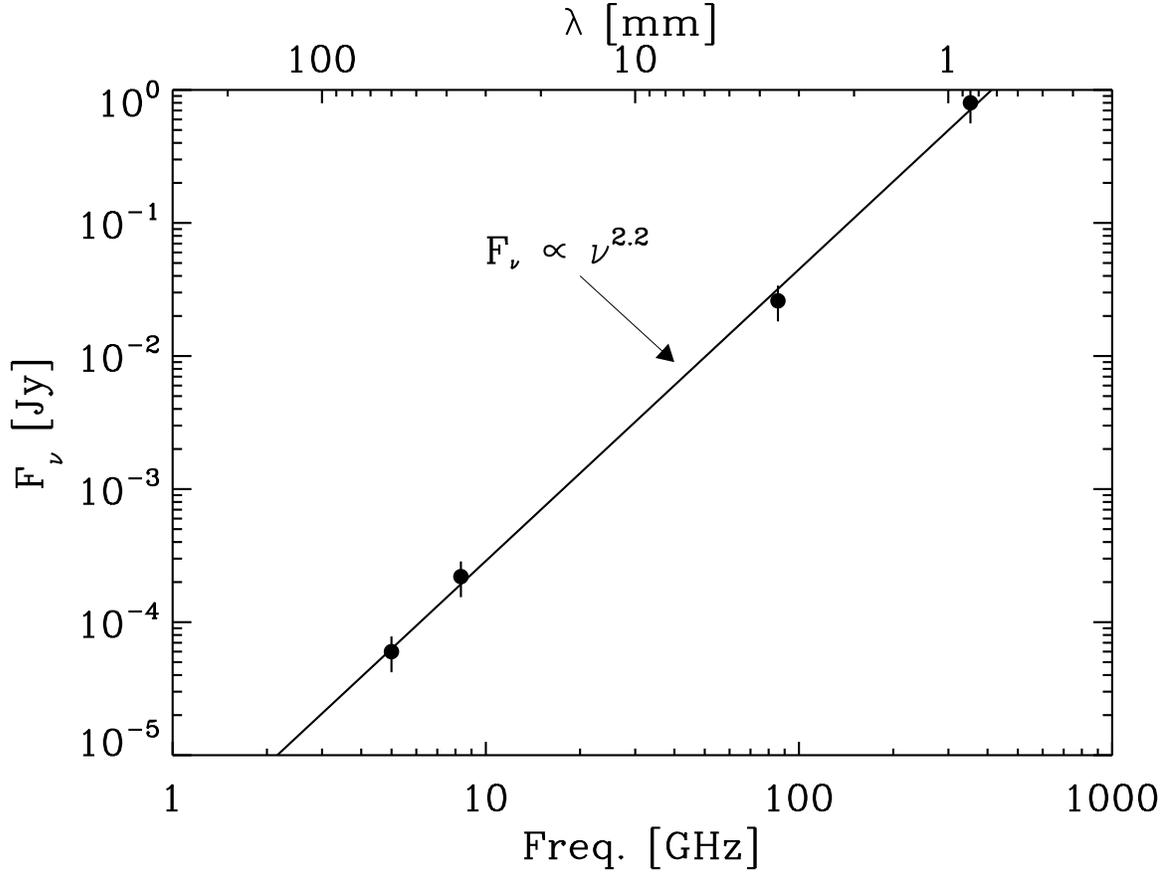}
\caption{Fluxes of disk component vs. frequency with cm observations
  from \cite{rodriguez99} and \cite{reipurth02}, 3.5~mm flux from the
  disk in an envelope with a $T_{\rm in}=75$~K cavity from
  \cite{n1333i2art} and likewise for the 850~$\mu$m observations from
  this paper.}\label{cont_specindex}
\end{figure}

These estimates of the disk column densities are simplistic and most
likely a lower limit to the dust mass. First, the temperature of the
dust may not be coupled to that of the gas. Lowering the dust
temperature, however, increases the dust mass required to produce the
same flux at the given wavelength. For example, adopting a disk
temperature of 30~K increases the dust mass by a factor 5 to
0.1~$M_\odot$. Second, Fig.~\ref{cont_specindex} illustrates that the
flux is consistent with a spectral index $\alpha= 2.2$ ($F_\nu \propto
\nu^\alpha$) from cm \citep{rodriguez99,reipurth02} through mm
\citep{n1333i2art} and submm wavelengths (this paper). The data in
Fig.~\ref{cont_specindex} are all from interferometer observations and
cover an unprecendented range up to 350~GHz. The power-law exponent of
$\approx$ 2 suggests that the compact source is caused by optically
thick thermal dust emission from dust with opacity $\kappa_\nu \propto
\nu$. It would therefore be more appropriate to estimate the dust mass
on the basis of the longer wavelength points. Taking the 3.6~cm
measurements of \cite{reipurth02}, rather than the 850~$\mu$m
measurements from this paper, results in a dust mass of
0.1--0.5~$M_\odot$ (for temperatures of 150--30~K). Of course at the
longer wavelengths it is less clear whether the dust opacity law is
valid and if sources other than dust could be contributing to the
observed compact emission.

As discussed in Sect.~\ref{continstruct}, the continuum interferometer
data resolve the compact structure into at least two components, which
we have modeled as an extended Gaussian structure (diameter of 300~AU)
and an unresolved component. The size of the Gaussian component is
similar to that of disks around more evolved T Tauri stars
\citep[e.g.,][]{kitamura02}. Similarly-sized disks are found around
other deeply embedded protostars \citep[e.g., L1157;][]{gueth97} but
smaller disks are found around other protostars \citep[e.g.,
  IRAS~16293-2422B;][]{rodriguez05}. An interesting question is
whether the smaller disks result from truncation in binary systems
\citep[e.g.,][]{artymowicz94} or whether there is an evolution of disk
size from the earliest stages and whether the smaller disks simply
reflect the youngest systems \citep[e.g.,][]{terebey84}. The fairly
extended Gaussian component in NGC~1333-IRAS2A would argue in favor of
the first suggestion, i.e., that a circumstellar disk quickly builds
up around low-mass protostars unless truncated in a binary
system. \cite{reipurth02} did not resolve the compact continuum source
seen at 3.6~cm with a beam size of about 0.3$''$ (66~AU). If the
3.6~cm source is related to the submillimeter/millimeter source as the
SED suggests, this would imply that the centimeter observations only
pick up the peak, unresolved component of the submillimeter
structure. The sensitivity of the centimeter observations by
\citeauthor{reipurth02} suggests that they would not detect an
extended (300~AU), lower surface brightness component such as seen in
the submillimeter observations.

In fact NGC~1333-IRAS2A may be a close ($< 65$~AU) binary itself such
as suggested by its quadrupolar outflow probed down to small scales
\citep[e.g.][]{n1333i2art}. It is possible that the compact component
seen in these data traces the disks around each of these binary
components with the resolved Gaussian component reflecting an extended
circumbinary disk. The unresolved source is responsible for about 30\%
of the total flux of the compact component suggesting that it contains
a similar fraction of the mass.

\subsection{Molecular emission from the circumstellar disk?}
In the discussion of the hot core scenario in \S\ref{lineabund} we
assumed that the envelope extends to small radii (23~AU; 0.1$''$)
where the temperature increases to 250~K. The analysis of the
continuum emission indicates that a disk is present around
NGC~1333-IRAS2, however, and an open question remains how much, if
any, of the observed line emission can be attributed to such a
disk. In particular, this is relevant if an inner cavity exists in the
envelope around NGC~1333-IRAS2: as shown in Fig.~\ref{cont_uvamp} the
continuum data are also consistent with an envelope where the
temperature never exceeds 75~K. Such an inner cavity might be expected
if the extended Gaussian component discussed above reflects the disk
formed as a result of the rotation of the core. The size of the cavity
(radius of 285~AU for the example with an envelope with an inner
temperature of 75~K) would then be the centrifugal radius of the
envelope within which the disk is formed
\citep[e.g.,][]{terebey84,stahler94,basu98}.

In this scenario the disk would also have to be warm as indicated by
the observed high excitation lines. Observations of compact CH$_3$OH
emission around another class 0 object, L1157, have suggested that the
CH$_3$OH emission is related to an accretion shock close to the
central disk \citep{goldsmith99,velusamy02}. \cite{ceccarelli02}
suggest that a disk with a heated accretion layer with a temperature
of 170--250~K exists around the class I young stellar object Elias 29
from high $J$ FIR/sub millimeter CO observations.
\clearpage
\begin{table}
\caption{Molecular abundances inferred from combined interferometer and single-dish data.}\label{abundstruct}
\begin{tabular}{lll|ll} \tableline\tableline
Molecule & \multicolumn{2}{c|}{Outer envelope}   & Envelope w. $R_{\rm i} = 23.4$~AU & Envelope w. $R_{\rm i} = 285$~AU \\
         & $X_0$$^{a}$ & $X_{\rm D}$$^{b}$  & Hot core abundance, $X_{\rm J}$$^{c}$ & Disk abundance, $X_{\rm disk}$$^{d}$ \\ \tableline
HCN$^{e}$                     &    2\tpt{-8}     & 2\tpt{-9}   &    7\tpt{-8}   &  2\tpt{-9}    \\
SO                            &    3\tpt{-9}     & $\ldots$    & $<$8\tpt{-9}   &  5\tpt{-10}   \\
SO$_2$                        & $<$2.5\tpt{-10}  & $\ldots$    &    2\tpt{-8}   &  9\tpt{-10}   \\
CH$_3$OH                      &    1.0\tpt{-9}   & $\ldots$    &    2\tpt{-6}   &  3\tpt{-8}    \\
CH$_3$OCH$_3$                 &    $\ldots$      & $\ldots$    &    3\tpt{-8}   &  2\tpt{-9}    \\ \tableline
\end{tabular}

$^{a}${Envelope abundance where $T < 90$~K (for HCN: where
furthermore $n_{\rm H_2}<7\times 10^4$~cm$^{-3}$). SO$_2$ and
CH$_3$OH abundances from single-dish observations by
\cite{paperii,hotcoresample}.} \\
$^{b}${Envelope abundance where $n_{\rm H_2}>7\times 10^4$~cm$^{-3}$ and $T < 90$~K.}  \\
$^{c}${Envelope abundance where $T > 90$~K in model extending to $T_i=250$~K (inner radius of 23.4~AU).} \\
$^{d}${Disk abundance in model where envelope extends to $T_i=75$~K (inner radius of 285~AU).} \\
$^{e}${Calculated from H$^{13}$CN abundance adopting $^{12}$C:$^{13}$C isotope ratio of 70.} \\
\end{table}
\clearpage
Assuming that the compact line emission comes from a homogeneous
medium with a density of $1\times 10^{9}$~cm$^{-3}$ (constant density
in a 100~AU thick disk given the estimated column density) and a
temperature of 150~K (see above), an estimate of the column density of
each molecular species can be made using the non-LTE escape
probability code \emph{Radex}\footnote{An online version of the code
  can be found at {\tt http://www.strw.leidenuniv.nl/$\sim$moldata}}
\citep{jansen94,schoeier03radex}. These can then be compared to the
column densities inferred from the dust emission above.  With these
assumptions, the SO and SO$_2$ abundances would be 5$\times 10^{-10}$
and 9$\times 10^{-10}$, respectively, and the H$^{13}$CN abundance
3$\times$10$^{-11}$ (corresponding to an HCN abundance of 2$\times
10^{-9}$ assuming standard isotopic ratios). Table~\ref{abundstruct}
summarizes the derived abundance structures. The value for the SO
abundance is consistent with the upper limit by \cite{thi04} toward
the pre-main sequence disks around LkCa15. The disk column densities
should be considered as lower limits (see Sect.~\ref{continorigin})
and therefore any abundances derived using the dust column densities
are upper limits. A more detailed treatment of the disk physical
structure (including its temperature and density variations) together
with more observations of the emission from a larger number of
molecules is needed for a complete picture of the disk chemistry in
these early deeply embedded stages.

Possibly the best proof for either the ``hot core'' or ``disk''
scenario can come from resolving the molecular emission and addressing
their intrinsic velocity fields. \cite{bottinelli04iras16293}
suggested that the discovery of organic molecules was evidence for hot
cores due to the alignment with the dust emission. This is, however,
not the case -- as their dust emission observed at smaller scales more
likely arises in the disks around the two binary components in
IRAS~16293--2422. In fact, if the reservoirs of organic molecules
observed in IRAS~16293--2422 are related to the larger scale envelopes
seen for example by single-dish continuum observations, molecular
species should be seen on intermediate scales between the ``compact
disk sources'' and the ``larger scale envelope''. Observations of more
complex, asymmetric sources such as IRAS~16293--2422 might therefore
be the best way to distinguish between the ``accretion disk'' and
``hot core envelope'' scenarios.

\section{Summary}
We have presented a detailed analysis of the physical and chemical
structure of the low-mass protostar, NGC~1333-IRAS2A based on high
angular resolution (1-2$''$; 200-400~AU) observations with the
Submillimeter Array. The data are compared to detailed continuum and
line radiative transfer models to zoom in on the inner few hundred AU
unhindered by emission from the lower density outer envelope. The
conclusions are as follows:

\begin{itemize}
\item Compact continuum emission is detected at 850$\mu$m: the SMA
  observations resolve the compact continuum component, which can be
  fitted by two structures in addition to the large scale envelope
  constrained by single-dish observations: A resolved, 300~AU sized,
  Gaussian brightness distribution and an unresolved ($< 200$~AU)
  structure, the latter contributing about 30\% of the flux of the
  compact continuum component.
\item The spectrum of the compact continuum component follows a
  power-law, $F\propto \nu^{2.2}$, from centimeter through
  submillimeter wavelengths. The favored explanation is that its
  origin is optically thick thermal dust emission from a circumstellar
  disk. The size of the disk around this young object suggests that
  the build-up of circumstellar disks proceeds rapidly in the
  protostellar stages. Smaller disks inferred around other protostars
  may be a consequence of tidal truncation by nearby binary
  companions.
\item Compact emission from lines of complex organic molecules
  including CH$_3$OCH$_3$ and CH$_3$OCHO, high excitation CH$_3$OH
  transitions, deuterated methanol CH$_3$OD as well as lines of CO,
  HCN, H$^{13}$CN, SO and SO$_2$ are detected. The line data are
  interpreted in the context of previously published single-dish
  studies of the chemical structure of the protostellar envelope. The
  SO data are consistent with a low constant abundance throughout the
  envelope - an SO abundance enhancement at small scales is ruled out
  by the current observations. The complex organic species and SO$_2$
  in contrast are enhanced on small scales. H$^{13}$CN follows the
  drop abundance structure also traced by CO and HCO$^+$ with a high
  abundance in the outer region where the density is low and no
  depletion has occurred, an intermediate region with low temperatures
  and higher densities where depletion can occur and an inner region
  where HCN either evaporates off dust grains due to high temperatures
  or resides in the circumstellar disk.
\item A disk of the size suggested by the continuum emission cannot be
  neglected in the interpretation of line data. If this disk has a
  warm ($\sim$100~K) layer, this could be the origin of the high
  excitation tracers and complex organic species on the smallest
  scales.
\end{itemize}

This paper illustrates the potential of the SMA for studying the
detailed variations in the physical and chemical structure of low-mass
protostellar envelopes, probing their innermost regions and
constraining the properties of circumstellar disks. The SMA data and
presented analysis will form a template for a large survey of deeply
embedded low-mass protostars currently ongoing at the SMA.

\acknowledgments We are grateful to the SMA staff and in particular
Charlie Qi for useful discussions about reduction of SMA data. We
extend special thanks to those of Hawaiian ancestry on whose sacred
mountain we are privileged to be guests. We thank the referee for a
prompt and well-considered report. This research was supported by NASA
Origins of Solar Systems Grant NAG5-13050. FLS acknowledges financial
support from the Swedish Research Council. Astrochemistry research in
Leiden is supported by a NWO Spinoza grant and a NOVA grant.

\end{document}